\documentclass[a4paper,fleqn,usenatbib]{mnras}

 \usepackage[normalem]{ulem}
 \usepackage[final]{pdfpages}
 
\usepackage[T1]{fontenc}
\usepackage{ae,aecompl}

\usepackage{graphicx}	
\usepackage{amsmath}	
\usepackage{amssymb}	
\usepackage{multirow}
\usepackage{multicol}
\usepackage{longtable}
\usepackage{supertabular}
\usepackage{natbib}
\usepackage{hyperref}

\newcommand{\emm}[1]{\ensuremath{#1}} 
\newcommand{\emr}[1]{\emm{\mathrm{#1}}}
 
\newcommand{\unit}[1]{\emr{\,#1}}

\newcommand{\kms}{\unit{km\,s^{-1}}} 
 
\newcommand{\K}{\unit{K}}

\newcommand{\pscm}{\unit{cm^{-2}}}

\title[Chemistry of HOCO$^+$]{Detection of HOCO$^+$ in the protostar IRAS 16293-2422}

\author[Majumdar et al.]{
L. Majumdar$^{1}$\thanks{E-mail: liton.majumdar@jpl.nasa.gov; liton.icsp@gmail.com}, P. Gratier$^{2}$, V. Wakelam$^{2}$, E. Caux $^{3,4}$, K. Willacy$^{1}$, M. E. Ressler$^{1}$
\\ 
$^{1}$ Jet Propulsion Laboratory, California Institute of Technology, 4800 Oak Grove Drive, Pasadena, CA 91109, USA\\
$^{2}$ Laboratoire d'astrophysique de Bordeaux, Univ. Bordeaux, CNRS, B18N, allée Geoffroy Saint-Hilaire, 33615 Pessac, France\\
$^{3}$ Universit\'e de Toulouse, UPS-OMP, IRAP, Toulouse, France\\
$^{4}$  CNRS, IRAP, 9 Av. Colonel Roche, BP 44346, F-31028 Toulouse Cedex 4, France
}
\date{Accepted XXX. Received YYY; in original form ZZZ}

\pubyear{2017}

\begin{document}
\maketitle

\begin{abstract} 
The protonated form of CO$_2$, HOCO$^+$, is assumed to be an indirect tracer of CO$_2$ in the millimeter/submillimeter regime since CO$_2$ lacks a permanent 
dipole moment. Here, we report the detection of two rotational emission lines (4$_{0,4}$--3$_{0,3}$ and 5$_{0,5}$--4$_{0,4}$) of HOCO$^+$ in IRAS 16293-2422. 
For our observations, we have used EMIR heterodyne 3 mm receiver of the IRAM 30m telescope. The observed abundance of HOCO$^+$ is compared 
with the simulations using the 3-phase NAUTILUS chemical model. Implications of the measured abundances of HOCO$^+$ to study the chemistry of CO$_2$ ices using JWST-MIRI and NIRSpec are discussed as well.

\end{abstract}




\section{Introduction}
Dust grains in the interstellar medium (ISM) are coated mostly with H$_2$O, CO, CO$_2$ ices along with other minor ice constituents formed in particular by surface chemistry. In star forming regions, these ices account for up to 60 and 80 percent of the volatile oxygen and 
carbon budget \citep{2011ApJ...740..109O}. CO$_2$ is an important constituent of these ices. CO$_2$ ice has been observed in dense clouds (\citet{1998ApJ...498L.159W}; \citet{2005ApJ...627L..33B}; \citet{2005ApJ...635L.145K}; 
\citet{2009ApJ...695...94W}; \citet{2013ApJ...775...85N}), protostellar envelopes (\citet{2013ApJ...775...85N};  \citet{2004ApJS..154..359B}; \citet{2008ApJ...678.1005P}; \citet{2010A&A...514A..12S}; \citet{2012A&A...538A..57A}) and in 
comets \citep{2012ApJ...752...15O}, the remnants of the proto-Solar Nebula. Irrespective of the different astrophysical sources, its abundance with respect to H$_2$O is constant around 20-30\%, which is one of the biggest puzzle for the astronomers (see below). 

In the literature, many plausible scenarios have been proposed to explain the formation of CO$_2$ ice in the ISM \citep{2008ApJ...678.1005P}.  \citet{1985A&A...152..130D} have suggested that strong UV irradiation is needed to produce the observed CO$_2$ ice 
since the reaction CO + O$\rightarrow$CO$_2$ on the surface has a large activation barrier. Later, \citet{1986A&A...158..119D} confirmed from laboratory experiment that CO$_2$ is formed from the ice mixtures of H$_2$O and CO under strong UV photolysis. 
The detection of CO$_2$ ice around UV-luminous massive young stars has confirmed this laboratory experiment (\cite{2008ApJ...678.1005P} and reference therein). But the detection of CO$_2$ ice in dark clouds with similar abundances (\citet{2005ApJ...627L..33B};
 \citet{2005ApJ...635L.145K}) has questioned the UV irradiation route to CO$_2$ since these sources are far away from any ionizing source. Later, with the help of a laboratory experiment, \citet{2001ApJ...555L..61R} claimed that the barrier for the oxygenation of CO is much lower than 
 previously assumed. Theoretical calculations and laboratory experiments are still a very active topic to understand the formation of CO$_2$ ice in the ISM \citep{2016ApJ...832....5C}.
 
 Through chemical models, CO$_2$ is also predicted to be one of the more abundant carbon and oxygen bearing molecules in the gas phase (\citet{1989ApJS...69..271H}; \citet{1991A&AS...87..585M}). In the solid phase, the observed abundance for CO$_2$ is of the 
 order of $10^{-5}$ to $10^{-6}$ relative to H$_2$ \citep{1999ApJ...522..357G}, e.g. a factor of 10 to 100 higher than in the gas phase (\citet{1996A&A...315L.349V}; \citet{2003A&A...399.1063B}). The large amount of CO$_2$ in the gas phase could be produced by 
 the evaporation or destruction of the icy grain mantles \citep{1996A&A...315L.349V}. Comparing the observed gas and ice phase CO$_2$ abundance ratio with those of other species (e.g. H$_2$O; CO) known to be abundant in icy mantles, will allow us to understand the 
 formation mechanisms of CO$_2$. 
 
 Carbon dioxide cannot be observed in the gas phase through rotational transitions in the far-infrared or submillimeter range due to its lack of permanent dipole moment. It has to 
 be observed through its vibrational transitions at near- and mid-infrared wavelengths. The protonated form of CO$_2$, HOCO$^+$ , is an interesting alternative to track the gas phase CO$_2$ in the millimeter/submillimeter regime. According to \citet{1977ApJ...215..503H},  
 the abundance of gas phase CO$_2$ relative to CO might be constrained by comparing the abundance of HOCO$^+$ with that of HCO$^+$. HOCO$^+$ was first detected in Galactic centre cloud SgrB2 \citep{1981ApJ...246L..41T} and later towards 
 SgrA \citep{1991A&A...244..470M}. HOCO$^+$ was also detected in low-mass Class 0 protostar
 IRAS 04368+2557 in L1527 \citep{2008ApJ...675L..89S}; in the prototypical protostellar bow shock of L1157-B1 \citep{2014A&A...565A..64P}; recently in L1544 prestellar core \citep{2016A&A...591L...2V}. 
 
  In this paper, we report the first detection of HOCO$^+$ in the low mass protostar IRAS 16293-2422 (hereafter IRAS 16293) and discuss its astrochemical implications as an indirect tracer of gas phase CO$_2$. 
 
 \section{Observations and data reduction}

\subsection{Observations}
The observations were performed at the IRAM-30m towards the midway point between sources A and B of IRAS 16293 at $\alpha_{2000} = 16^h32^m22.75^s, \delta_{2000} = -24\degr28\arcmin34.2\arcsec$. We preformed our observations during the 
period of August 18 to August 23, 2015 under average summer conditions (a median value of 4-6 mm water vapour). For our observations, we have used the EMIR heterodyne 3 mm receiver tuned to a frequency of 89.98 GHz 
in the Lower Inner sideband, connected to the FTS spectrometer in its 195 kHz resolution mode. Our observed spectra was composed of two regions: one from 84.4 GHz to 92.3 GHz and another one from 101.6 GHz to 107.9 GHz.

\begin{figure*}
   \includegraphics[width=\textwidth]{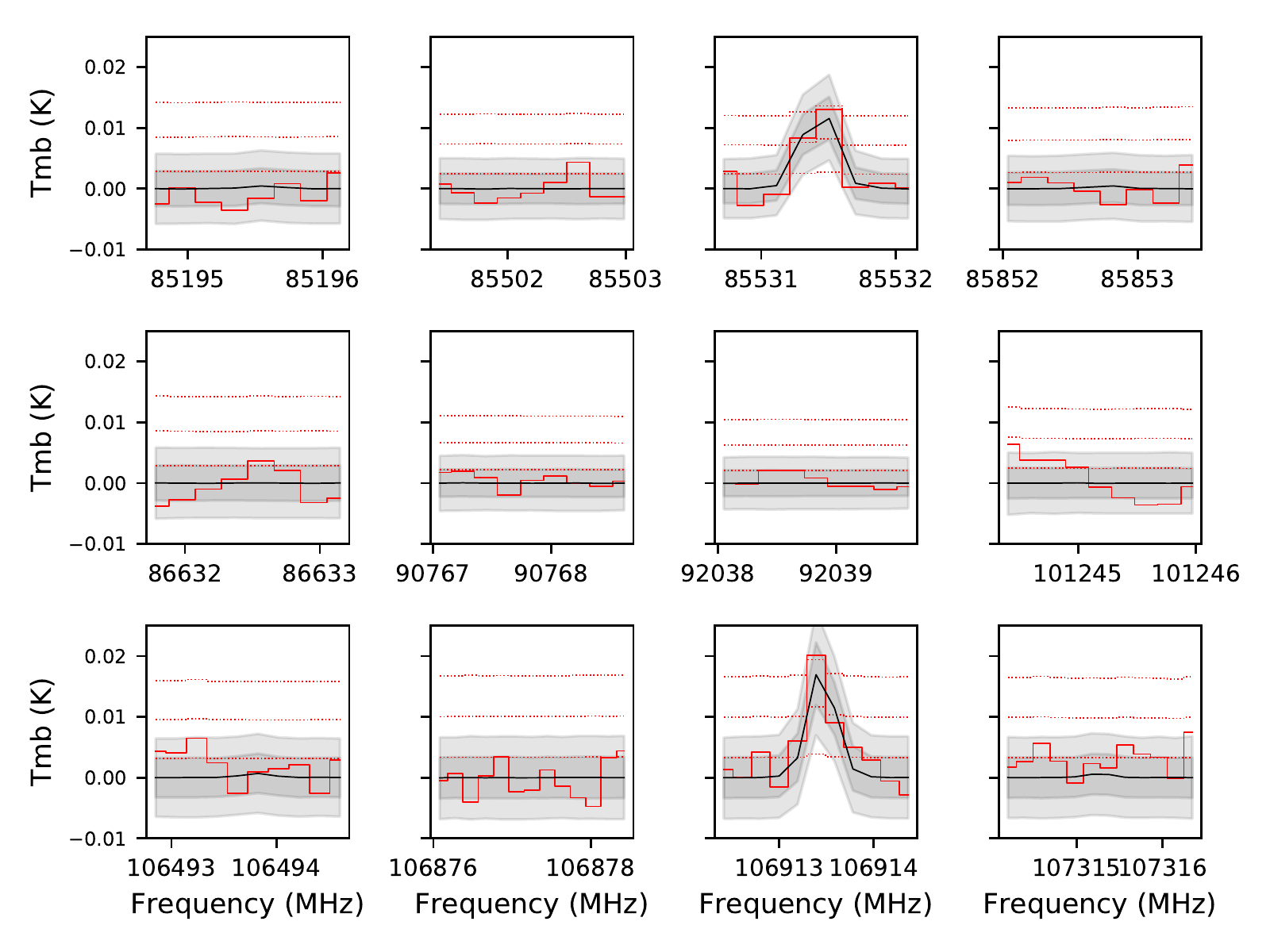}
    \caption{\label{fig.gauss_fit} The red lines show the observed lines attributed to HOCO$^+$. 
    The black lines show the distribution of modelled spectra following the posterior distribution of parameters shown in Fig. 2. The thick line denotes the median 
    of the distribution. The dark and light grey regions show the  68\% and 95\% confidence intervals. The dotted lines are 1$\sigma$, 3$\sigma$ and 5$\sigma$ noise levels. 1$\sigma$ level is 2.7 mK.}
\end{figure*}

\subsection{Results}

\begin{table*} 
\caption{Observed lines and spectroscopic parameters{\bf $^\ddagger$} for HOCO$^+$} 
\begin{center}
 \begin{tabular}{cccccccc}

Transitions  &  Frequency &        Aij        &   E$_{up}$  &  V$_{LSR}$       &  FWHM              &  Integrated flux   \\

                   &     (MHz)      &   (s$^{-1}$) &   (K)          &  (km~s$^{-1}$)  & (km~s$^{-1}$)     &     (K~km~s$^{-1}$) \\
\hline
 4$_{0,4}$--3$_{0,3}$  & 85531.497  & 2.36$\times$10$^{-5}$ & 10.3  & 4.06 $\pm$ 0.11  & 0.69 $\pm$ 0.03  & 0.015 $\pm$ 0.004 \\
5$_{0,5}$--4$_{0,4}$  & 106913.545 & 4.71$\times$10$^{-5}$ &15.4  & 4.16 $\pm$ 0.09  & 0.94 $\pm$ 0.04  & 0.020 $\pm$ 0.004 \\
\hline
\end{tabular}
\begin{minipage}{11cm} {$^\ddagger$Spectroscopic data have been taken from \citet{2017A&A...602A..34B}} \end{minipage}   
 \end{center}
\end{table*} 

\subsubsection{HOCO$^+$ line properties}
We have used the CLASS software from the GILDAS\footnote{\url{https://www.iram.fr/IRAMFR/GILDAS/}} package for our data reduction and analysis. 
We did Gaussian fits to the detected lines after subtracting a local low (0 or 1) order polynomial baseline subtraction. In Table 1, we have shown the result of these fits 
for the two detected lines of HOCO$^+$. Both these detected lines are single-component features with mean LSR velocity of $\sim$ 4.1 km/s and a mean 
FWHM of $\sim$ 0.81 km/s. In the past, \citet{2011A&A...532A..23C} has defined four types of kinematical behaviours of IRAS 16293 for various species based on their 
FWHM and V$_{LSR}$ distributions. Here, we find that HOCO$^+$ belongs to the type I (i.e. FWHM$\leq$2.5 km/s, VLSR$\sim$ 4 km/s, Eup$\sim$ 0-50 K) and this
corresponds to species abundant in the cold envelope.

\begin{figure*}
    \includegraphics[width=\textwidth]{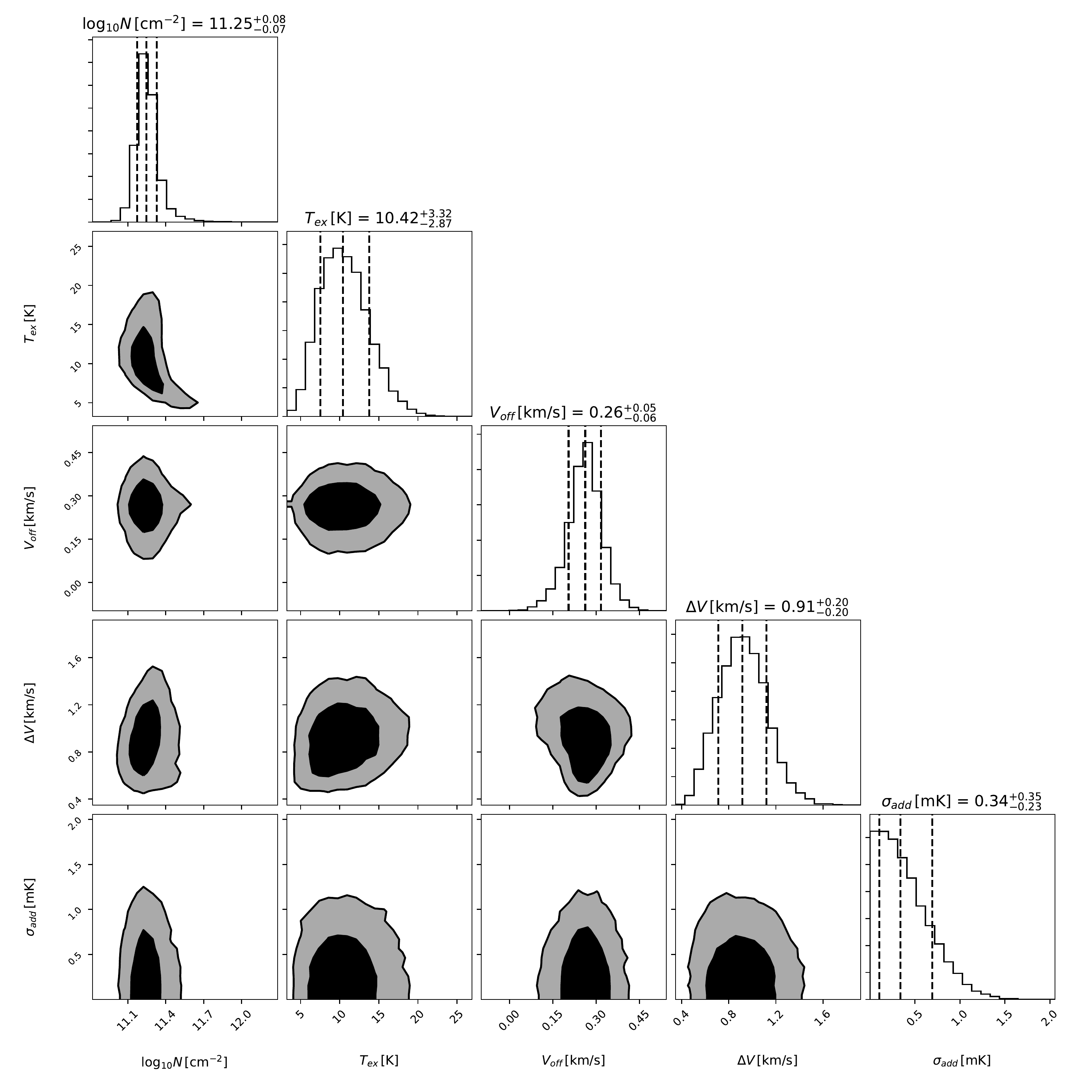}
       \caption{\label{fig.bayes_result} 1D and 2D histograms of the
   posterior distribution of parameters for HOCO$^+$. Contours contain 68 and 95 \% of samples, respectively. Uncertainty quoted here are statistical only, without 10\% calibration error. $V_{off}$ is the difference between the observed line position and the reference $V_{lsr}$ position for IRAS 16293 of 3.6 km/s ($V_{line}$ = $V_{lsr}$ + $V_{off}$ with $V_{lsr}$=3.8 km/s).} 
    \end{figure*}

\begin{figure*}
  \includegraphics[width=150mm]{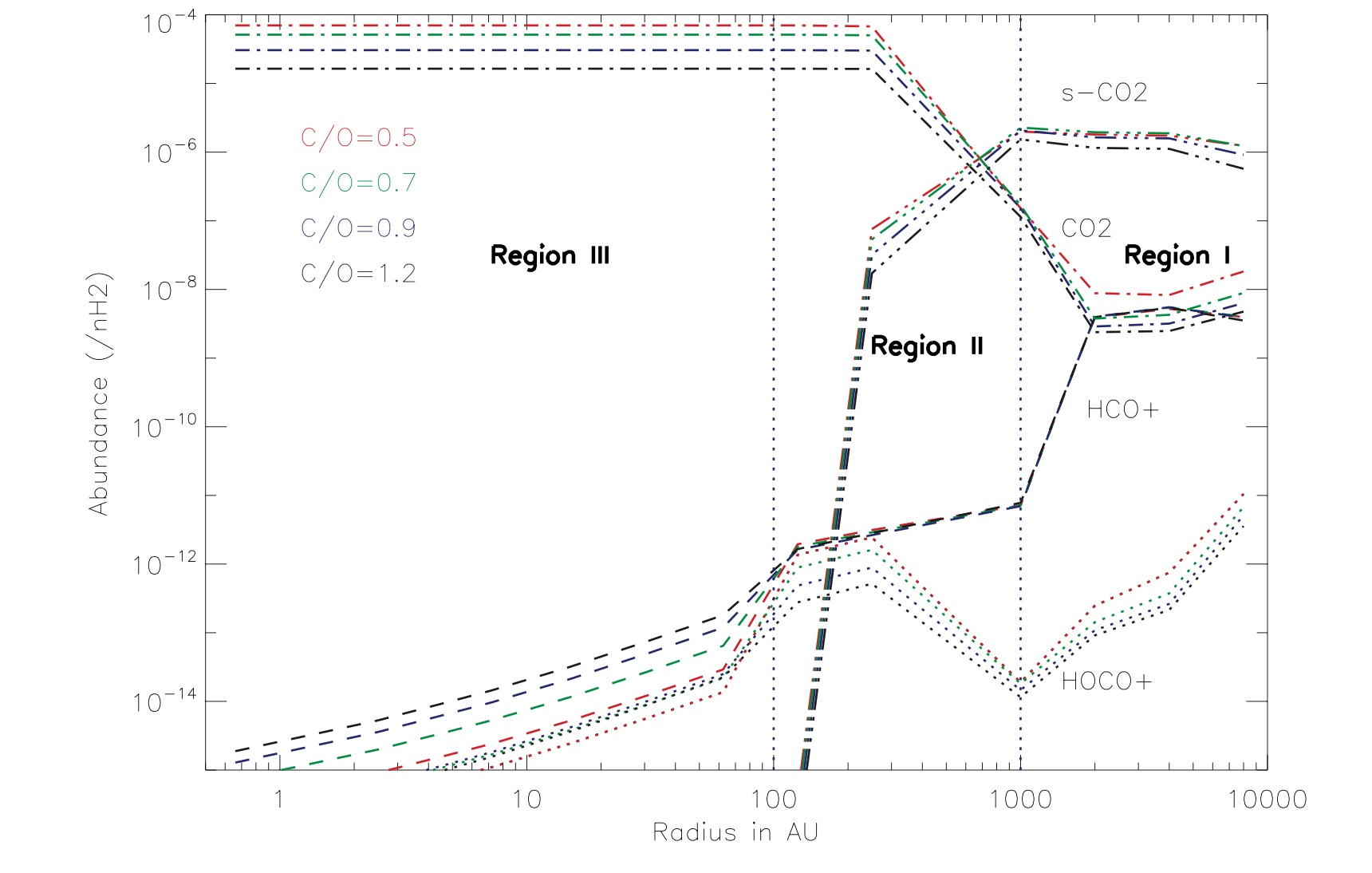}
  \caption{Abundance with respect to H$_2$ for HOCO$^+$, HCO$^+$, CO$_2$ predicted by our model as a
function of radius. s-CO$_2$ represents the CO$_2$ on the surface of grains.}
\end{figure*}

\subsubsection{LTE modelling of HOCO$^+$}

We have used the bayesian model similar to the one used in \citet{2017MNRAS.467.3525M}. They used this model to recover the distribution of parameters which best agree 
with the observed line intensities of {\it c}-C$_3$H$_2$ and {\it c}-C$_3$HD in the same source. 

In order to model the emission of HOCO$^+$, we have used an LTE radiative transfer code based on the equations described in \citet{2011A&A...526A..47M}. The model needs 
the species column density, the line width, the excitation temperature, the source size and an accurate molecular spectroscopic catalog which contains energy levels with associated quantum numbers, statistical weights and transition frequencies 
as well as integrated intensities at 300 K.  For HOCO$^+$, we have used the spectroscopic data from \citet{2017A&A...602A..34B} retrieved from the CDMS database \citep{2005JMoSt.742..215M}. All the detected frequencies along with their Einstein coefficients, upper level energies and the 
associated quantum numbers are listed in Table 1. 

Since we have only two detected lines, we have considered a fixed source size of 25 arcsec ($\sim$ 3000 AU, a typical size of the protostellar envelope; \citet{2011A&A...532A..23C}). In our model, the likelihood function assumes that the errors are 
distributed with a noise term.  This noise term is defined as the sum in quadrature of the observed per channel uncertainty and an additional noise term (noted as $\sigma_{add}$ in Figure 2) left as a free parameter in the model. By following \citet{2017MNRAS.467.3525M}, here also we 
have assumed similar distribution of the priors. We carried out sampling of the posterior distribution using the Hamilonian Monte Carlo NUTS sampler implemented in the Stan package \footnote{\url{http://mc-stan.org}} and the PyStan wrapper \footnote{\url{http://mc-stan.org/interfaces/pystan}}. The sampling was run for 2000 iterations using 6 independent chains, with the 1000 first iterations discarded for warmup. Convergence was checked by computing the split $\hat{R}$ estimator \citep{Gelman.1992} for all parameters, all of which were found to be less than 1.01.

Figure 1 shows the comparison of the observed and modelled spectra for HOCO$^+$. In Figure 2, we have shown the 1D and 2D histograms of the posterior 
probability distribution function for HOCO$^+$. From Figure 2, it is very clear that excitation temperature, systematic velocity and line width are well defined. In Table 2, we have summarised the 
one point statistics for the marginalised posterior distributions of parameters with 1$\sigma$ symmetric error bars in the parentheses. We have used the H$_2$ column density of 1.6$\times$ 10$^{24}$ cm$^{-2}$ derived by \citet{2002A&A...390.1001S} in the cold envelope to derive the HOCO$^+$ abundance. We have also listed modelled abundance of HOCO$^+$ at 3000 AU discussed in the Section 3. 

\begin{table}
    
    \caption{Point estimates of the posterior distribution function
    corresponding to the median and one sigma uncertainty.}
    
    \label{tab.bayes_result}
    \centering
    \begin{tabular}{lc}
        \hline
        Parameter           & Value  \\
        \hline
           \emr{\log N (HOCO^+)} (\pscm)   &   $11.25\pm0.08$         \\
           \emr {T_{ex}} (\K)      &   $10    \pm 3$         \\
           \emr{ \Delta V} (\kms)  &   $ 0.9\pm0.2$         \\
           \emr{\log [HOCO^+]^{observation}}    &   $-12.95\pm0.08$         \\
          \emr{\log [HOCO^+]^{model (C/O=0.5)}}    &     -12.32      \\     
                   
         \emr{\log [HOCO^+]^{model (C/O=0.7)}}    &     -12.61      \\     
         
         \emr{\log [HOCO^+]^{model (C/O=0.9)}}    &     -12.76      \\     
         
          \emr{\log [HOCO^+]^{model (C/O=1.2)}}    &     -12.85      \\

       \hline
                
    \end{tabular}
\end{table}

\section{Chemistry of HOCO$^+$ in the proto-stellar envelope}

\subsection{The NAUTILUS chemical model}
We have investigated the chemistry of HOCO$^+$ in IRAS 16293 by using the state-of-the-art {\tt NAUTILUS} three phase chemical model \citep{2016MNRAS.459.3756R}. {\tt NAUTILUS} computes the chemical
composition as a function of time in the gas-phase, and at the surface of interstellar grains. All the physicochemical processes included in the model along with their corresponding equations are  described in detail in 
\citet{2016MNRAS.459.3756R}. Our gas phase chemistry is based on the public chemical network kida.uva.2014 \citep{2015ApJS..217...20W}. The surface network is based on the one of \citet{2007A&A...467.1103G} with
several additional processes from \citet{2015MNRAS.447.4004R}. By following \citet{2011A&A...530A..61H}, we adopt the similar initial elemental abundances with an additional elemental abundance 
of $6.68\times10^{-9}$ for fluorine (\citet{2005ApJ...628..260N}) and different C/O elemental ratios of 0.5, 0.7, 0.9, 1.2. 

\subsection{1D physical model} 
The physical model for IRAS 16293 follows the one detailed in our previous papers (\citet{2016MNRAS.458.1859M}; \citet{2017MNRAS.467.3525M}). It was based on the radiation hydrodynamical (RHD) simulations from \citet{2000ApJ...531..350M}. This physical 
dynamical model starts from a dense molecular cloud with a  central density $n$(H$_2$) $\sim 3 \times 10^4$ cm$^{-3}$ and the core extends up to  $r=4 \times 10^4$ AU with a total mass of 3.852 $M_{\odot}$, which exceeds the critical mass for gravitational instability. The initial 
temperature for the core is around 7 K at the center and around 8 K at the outer edge. In order to set up the initial molecular conditions for the collapse stage, the core stays at its hydrostatic structure for 10$^6$ year. 

After 10$^6$ year, the contraction starts for the core and it is almost isothermal as long as the cooling is efficient. When compressional heating overwhelms the cooling, it causes a rise of the temperature in the central region. Eventually, the first hydrostatic core forms when 
contraction decelerates due to the increase of the gas pressure. This is also known as `first core' at the center. 

A second collapse happens when the core center becomes unstable due to a very high density of 10$^7$ cm$^{-3}$ and a high temperature of 2000 K 
which causes H$_2$ dissociation.  Within a short period of time, the dissociation degree approaches unity at the center due to the rapid increase of central density. Then the second collapse
ceases, and the second hydrostatic core, i.e., the protostar, is formed and the infalling envelope around this protostar is known as `protostellar core'. It takes a $2.5 \times 10^5$ yr for the initial pre-stellar core to evolve into the protostellar core. 

When the protostar is formed, the model again follows the evolution for a $9.3\times10^4$ yr, during which the protostar grows by mass accretion from the envelope.

\subsection{Results and discussions}
Figure 3 shows the computed abundance of HOCO$^+$ in the gas phase in the protostellar envelope as a function of radius to the central protostar, at the end of the simulations (i.e. at the protostellar age of $9.3\times10^4$ yr) for different C/O elemental ratios of 0.5, 
0.7, 0.9, 1.2. The ``age" that we considered for the protostar is set by the physical dynamical model. As discussed in \citet{2014MNRAS.445.2854W}, the physical structure obtained with the radiation hydrodynamical model at that time is similar to the one constrained by 
multi-wavelength dust and molecular observations (from \citet{2010A&A...519A..65C}). We have also shown the 
abundance profile of HCO$^+$ and CO$_2$ as they are the major precursors for the formation of HOCO$^+$. 

\citet{2011A&A...530A..61H} have already performed a detailed sensitivity analysis to the different oxygen elemental abundances. They have concluded that gas phase abundances calculated with the {\tt NAUTILUS} gas-grain chemical model are less sensitive to the 
elemental C/O ratio than those computed with a pure gas phase chemical model \citep{2010A&A...517A..21W}. This is consistent with our current results shown in Figure 3. The grain surface chemistry plays the role of a buffer absorbing most of the extra carbon. This is fortunate because 
the C depletion problem is still poorly understood from observations. This reduced sensitivity of the chemistry to the C/O ratio makes this conclusion more robust. 

For our discussion, we divide our model into three different regions: (i) a first region (radii larger than 1000 AU and temperatures below 30 K), (ii) a second region (radius in between 1000 and 200 AU and temperature in between 30 and 60 K) (iii) and a third region (radii lower than 200 AU and temperatures 
higher than 60 K; $\sim$ 70-250 K in between 150 to 10 AU). 

In the first region, HOCO$^+$ is mainly formed from the  OH + HCO$^+$ and CO$_2$ + H$_3$$^+$ reactions. The contribution of OH + HCO$^+$$\rightarrow$ H + HOCO$^+$ reaction is $\sim$85\% towards the HOCO$^+$ formation. The contribution of the second reaction is 
much less since CO$_2$ is frozen on the grains. In this region, HCO$^+$ is the main precursor for HOCO$^+$ formation. The HCO$^+$ is mainly formed from the CO + H$_3^+$ reaction whereas OH radical originates mostly from the dissociative recombination reactions of H$_3$O$^+$. Formation of 
H$_3$$^+$ and hence HCO$^+$ are governed by the reaction with cosmic rays. Thus, in the outer part of the envelope, the HOCO$^+$ abundance is controlled by the cosmic rays.

In the second region, HOCO$^+$ is mainly formed from the CO$_2$ + H$_3^+$ reaction. The contribution of this reaction is $\sim$85\% towards the HOCO$^+$ formation. In this region, CO$_2$ is the major precursor for HOCO$^+$ formation since the abundance of gas phase CO$_2$ starts increasing due to thermal desorption in the inner part of the envelope. A small fraction of HOCO$^+$ is also produced from the CO$_2$ + N$_2$H$^+$ reaction and the efficiency of the OH + HCO$^+$ reaction is almost negligible due to the rapid fall in HCO$^+$ abundance. Here, HOCO$^+$ is destroyed by CO + HOCO$^+$$\rightarrow$CO$_2$ + HCO$^+$  and CH$_4$ + HOCO$^+$$\rightarrow$CO$_2$ + CH$_5$$^+$ reactions. The second HOCO$^+$ peak around 200 AU is due to the effect of thermal desorption of the CO$_2$ ice. 

Recently, \cite{2016A&A...591L...2V} have observed HOCO$^+$ in the L1544 prestellar core. The observed abundance was ($5\pm 2 )\times10^{-11}$ with respect to molecular hydrogen. They also performed modeling of HOCO$^+$ in this source using a gas phase chemical code. In their model, they 
used a C/O ratio of 0.5, cosmic ray ionization rate of $3\times10^{-17}$ s$^{-1}$, n(H) equal to $2\times10^4$ cm$^{-3}$, a temperature of 10 K and A$_v$=10 magnitude. Finally, they compared the steady-state abundance of HOCO$^+$ which is equal to $4\times10^{-11}$ with respect to molecular hydrogen
with the observed abundance. The main conclusion from their model is that the chemistry of HOCO$^+$ depends on the reaction HCO$^+$ + OH$\rightarrow$HOCO$^+$ + H when CO$_2$ is frozen on the grains.  HOCO$^+$ abundance depends on the CO$_2$ + H$_3^+$$\rightarrow$HOCO$^+$ + H$_2$ reaction when gaseous CO$_2$ abundance is increased due to desorption processes. The major reactions that contributed to forming HOCO$^+$ in \cite{2016A&A...591L...2V} are consistent with our findings. The only exception is cosmic-ray induced UV photo-desorption \citep{2012ApJ...759L..37C} is likely for 
CO$_2$ ice in L1544 as compared to thermal desorption \citep{2016A&A...591L...2V}.

The HOCO$^+$ and CO$_2$ abundance profile in the gas phase are correlated till 200 AU.  In the third region (radii lower than 200 AU), HOCO$^+$ and HCO$^+$ abundances decrease rapidly due to very efficient destruction channels via H$_2$O + HOCO$^+$$\rightarrow$ CO$_2$ + H$_3$O$^+$ ($\sim$ 77\%), CO + HOCO$^+$$\rightarrow$ CO$_2$ + HCO$^+$ ($\sim$ 15\%), H$_2$O + HCO$^+$$\rightarrow$ CO + H$_3$O$^+$ 
($\sim$ 67\%) and HCN + HCO$^+$$\rightarrow$ CO + HCNH$^+$ ($\sim$ 18\%) reactions.  Due to close interplay of these four reactions, the abundance of HOCO$^+$ is correlated with that of HCO$^+$ in the inner part of the envelope.

The abundances of HOCO$^+$ predicted by our model at 3000 AU (approximate size of the envelope) are $4.7\times10^{-13}$, $2.4\times10^{-13}$, $1.7\times10^{-13}$ and $1.4\times10^{-13}$ with respect to molecular hydrogen for C/O ratios of 0.5, 0.7, 0.9 and 1.2 respectively. 
Looking at the small differences between these chemical models with different C/O ratios, we think HOCO$^+$ is not a good tracer to constrain the elemental C/O ratios in IRAS 16293. The observed abundance of HOCO$^+$ is ($1.12\pm 0.3 )\times10^{-13}$ with respect to 
molecular hydrogen (See Table 2). We have used the H$_2$ column density of 1.6$\times$ 10$^{24}$ cm$^{-2}$ from \citet{2002A&A...390.1001S}
 which corresponds to the full line of sight including the hot corino whereas HOCO$^+$ was assumed to be coming from the envelope. Therefore the derived observed abundance is most probably a lower limit to the abundance in the outer envelope. 
 \citet{2002A&A...390.1001S} have also derived the abundance of HCO$^+$ in the cold envelope by assuming the same H$_2$ column density of 1.6$\times$ 10$^{24}$ cm$^{-2}$  and the derived value was $1.4\times10^{-09}$ with respect to molecular hydrogen. We compare the abundance ratios 
 of [HOCO$^+$] to [HCO$^+$] which yield a value of 8$\times$ 10$^{-5}$ from the observation. This value is reasonably close to the ratio 5$\times$$10^{-5}$ from our model (by considering [HOCO$^+$]$\sim$ $2.4\times10^{-13}$ and [HCO$^+$]$\sim$ $4.8\times10^{-09}$ for a standard C/O ratio of 0.7 
 \citep{2011A&A...530A..61H} used generally in our models). \cite{2016A&A...591L...2V} have derived  [HOCO$^+$] to [HCO$^+$] ratio in  L1544 by assuming the bulk of the emission arrises from the outer layer. The derived value was 6.2$\times$$10^{-3}$ by considering an H$_2$ column density of 
 5$\times$ 10$^{21}$ cm$^{-2}$. Comparative studies with higher spatial resolution using ALMA/NOEMA interferometers may help us to understand the underlying physics and chemistry better. 

At 3000 AU, the abundance of CO$_2$ in the ice phase is $2\times10^{-6}$ with respect to molecular hydrogen. Higher abundance of CO$_2$ ice in the protostellar envelope shows that the JWST-MIRI and  NIRSpec will be able to give a wealth of information on the chemistry of 
CO$_2$ formation on molecular ices. With James Webb Space Telescope (JWST), scheduled to be launched in 2019, the spectroscopy of molecular ices could be done in the 0.6 to 28.5 micron wavelength range. It is approximately 100 times more powerful than Hubble and 
Spitzer space telescopes. It has greater sensitivity, higher spatial resolution in the infrared and significantly higher spectral resolution in the mid infrared. CO$_2$ ice has two strong vibrational modes; the asymmetric stretching mode centered on 4.27 micron, and the bending mode at 15.2 micron, 
accessible via JWST NIRSpec (0.6-5 micron with single slit spectroscopy mode) and MIRI (11.9-18 micron with IFU mode). These infrared vibrational features are sensitive to the astrophysical environments and band profiles can guide astronomers whether CO$_2$ molecules are embedded in H$_2$O, CO ices or other CO$_2$ isotopes \citep{2017A&A...601A..36B}. Detection of HOCO$^+$ in the solar type protostar IRAS 16293 will motivate future observation of CO$_2$ (which is one of the important constituent of planetary atmospheres) in similar type of environments using JWST. 

\section*{Acknowledgements}

This work is based on observations carried out with the IRAM
30m Telescope. IRAM is supported by INSU/CNRS (France), MPG (Germany) and
IGN (Spain). VW thanks ERC starting grant (3DICE, grant agreement 336474) for funding during this 
work. LM also acknowledges support  from  the  NASA  postdoctoral  program. VW and PG acknowledge the French program Physique
et Chimie du Milieu Interstellair (PCMI) funded by the Conseil
National de la Recherche Scientifique (CNRS) and Centre National
dEtudes Spatiales (CNES).  A portion of this research was carried out at the Jet Propulsion Laboratory, California Institute of Technology, under a
contract with the National Aeronautics and Space Administration. We would like to
thank the anonymous referee for constructive comments that helped
to improve the manuscript.


\bibliographystyle{mnras}
\bibliography{HOCO+}

\end{document}